\documentclass[english]{article}
\usepackage[T1]{fontenc}
\usepackage[latin1]{inputenc}
\usepackage{amsmath}
\usepackage{amssymb}
\usepackage{babel}

\makeatletter
\newtheorem{theorem}{Theorem}
\newtheorem{acknowledgement}[theorem]{Acknowledgement}

\input{tcilatex}
\makeatother

\begin{document}

\title{On the action principle for a system of differential equations}
\author{D.M. Gitman$^{1}$, V.G. Kupriyanov$^{1,2}$ \\
%EndAName
\\
$^{1}$Instituto de Física, Universidade de São Paulo, Brazil\\
$^{2}$ Physics Department, Tomsk State University, Tomsk\textit{,} Russia\\
E-mail: gitman@dfn.if.usp.br (D.M.Gitman), kvg@dfn.if.usp.br
(V.G.Kupriyanov).}
\maketitle

\begin{abstract}
We consider the problem of constructing an action functional for physical
systems whose classical equations of motion cannot be directly identified
with Euler-Lagrange equations for an action principle. Two ways of action
principle construction are presented. From simple consideration, we derive
necessary and sufficient conditions for the existence of a multiplier matrix
which can endow a prescribed set of second-order differential equations with
the structure of Euler-Lagrange equations. An explicit form of the action is
constructed in case if such a multiplier exists. If a given set of
differential equations cannot be derived from an action principle, one can
reformulate such a set in an equivalent first-order form which can always be
treated as the Euler-Lagrange equations of a certain action. We construct
such an action explicitly. There exists an ambiguity (not reduced to a total
time derivative) in associating a Lagrange function with a given set of
equations. We present a complete description of this ambiguity. The general
procedure is illustrated by several examples.
\end{abstract}

\section{Introduction}

The problem of constructing an action functional for a given set of
differential equations is known in literature as the inverse problem of the
calculus of variations for the Newtonian mechanics. In its classical setting 
\cite{H} consists of solving the variational equation 
\begin{equation}
\frac{\delta S[q]}{\delta q^{i}(t)}=g_{i}\,,  \label{f1}
\end{equation}
where $g_{i}(t,q^{i},\dot{q}^{i},...)=0$ is some given system of
differential equations with respect to unknown functions $q^{i}(t)$, and $%
S[q]$ is a local functional to be determined. The condition of locality
requires the existence of a function $L(t,q,\dot{q}...)$ (Lagrangian), such
that the functional $S[q]$ (action) would be written as an integral 
\begin{equation}
S[q]=\int dtL\,.
\end{equation}
In other words, the essence of the inverse problem of the calculus of
variations consists of finding a variational principle for a given system of
differential equations. This problem has been under consideration for more
than a hundred years. As early as 1887 Helmholtz \cite{H} presented a
criterion of commutativity for second variational derivatives from which
immediately follows the necessary (and with some restrictions also
sufficient) condition of solvability of the equation (\ref{f1}): 
\begin{equation}
\frac{\delta g_{i}(t)}{\delta q^{j}(s)}=\frac{\delta g_{j}(s)}{\delta
q^{i}(t)}\,.  \label{f3}
\end{equation}
If this condition holds, the system $g_{i}(t,q^{i},\dot{q}^{i},...)=0$ is
called Lagrangian system, if not the system is non-Lagrangian. In 1894
Darboux \cite{Darbu} solved the problem for the one dimensional case. In
1941 the case of two degrees of freedom was investigated by Douglas \cite%
{Duglas}; in particular, he presented examples of second-order equations
which cannot be obtained from the variational principle. Afterwards many
authors (see e.g., \cite{Sarlet1}-\cite{GK} and references therein)
investigated this problem for multidimensional systems.

In the present work we consider the question of the construction of an
action principle for a given system of differential equations using the
integrating multiplier method \cite{Duglas}-\cite{Sarlet}. The integrating
multiplier is a nonsingular matrix which being multiplied by a given set of
differential equations reduces this set to a standard Euler-Lagrange form.
In section 2 we present a simple derivation for necessary and sufficient
conditions for an integrating multiplier for a system of second-order
equations. We also construct the explicit form of the Lagrangian in case an
integrating multiplier exists. Then we apply our method for investigating
the inverse problem of some simple models. In particular, we construct an
action principle for multidimensional dissipative systems. We also consider
an example of a linear dynamical system whose equations of motion does not
admit an integrating multiplier, and, as a consequence, cannot be obtained
from the minimum action principle.

Note that it is always possible to reduce non-Lagrangian second-order
equations of motion to an equivalent set of first-order differential
equations. From the Helmholtz criterion (\ref{f3}) we find necessary and
sufficient conditions for the existence of an integrating multiplier for
such equations. It turns out that in the first-order formalism an
integrating multiplier always exists and can be constructed by means of the
solution of the Cauchy problem for the equations in question, and this is
presented in section 3 and is partially based on results of the works \cite%
{Henneaux} and \cite{GK}. Then we construct the action functional
explicitly. Thus, we show that systems traditionally called as
non-Lagrangian are, in fact, equivalent to some first-order Lagrangian
systems. As an example, we construct a first-order action functional for any
linear dynamical system.

\section{Action functional for a set of second-order equations}

\subsection{General consideration}

Let a system with $n$ degrees of freedom be described by a set of $n$
second-order differential equations of motion, solvable with respect to
second-order time derivatives. Suppose such a set has the form 
\begin{equation}
\ddot{q}^{i}-f^{i}(t,q,\dot{q})=0,\;\;\;\;\;\;\;\;\;\; i=1...n,  \label{1}
\end{equation}
where $f^{i}(t,q,\dot{q})$ are some functions of the indicated arguments,
and by dots above we denote time derivatives of the coordinates. Let us
construct an action principle for this set. If (\ref{1}) cannot be directly
identified with Euler-Lagrange equations, one can find an integrating
multiplier, i.e., a nonsingular matrix $\ h_{ij}(t,q,\dot{q})$ that being
multiplied by (\ref{1}) 
\begin{equation}
h_{ij}\left[\ddot{q}^{j}-f^{j}(t,q,\dot{q})\right]=0  \label{2}
\end{equation}
reduces this set to standard Euler-Lagrange form for some Lagrangian $L(t,q,%
\dot{q})$, 
\begin{equation}
\frac{\partial L}{\partial q^{i}}-\frac{\partial^{2}L}{\partial t\partial%
\dot{q}^{i}}-\frac{\partial^{2}L}{\partial\dot{q}^{i}\partial q^{j}}\dot{q}%
^{j}-\frac{\partial^{2}L}{\partial\dot{q}^{i}\partial\dot{q}^{j}}\ddot{q}%
^{j}=0\,.  \label{3}
\end{equation}
In order to identify (\ref{2}) with (\ref{3}) we need to ensure that 
\begin{eqnarray}
& & \frac{\partial^{2}L}{\partial\dot{q}^{i}\partial\dot{q}^{j}}%
\,=h_{ij},\;\;\;\;\;\;\;  \label{4} \\
& & \frac{\partial L}{\partial q^{i}}-\frac{\partial^{2}L}{\partial t\partial%
\dot{q}^{i}}-\frac{\partial^{2}L}{\partial\dot{q}^{i}\partial q^{j}}\dot{q}%
^{j}=h_{ij}f^{j}\,.\;\;\;  \label{5}
\end{eqnarray}
Provided that an integrating multiplier is known, equations (\ref{4})-(\ref%
{5}) can be interpreted as a system of equations for a Lagrange function $L$%
. We are going to solve the set of equations (\ref{4})-(\ref{5}). Its
consistency conditions will give us all the necessary and sufficient
conditions for an integrating multiplier. Assuming that $L$ is a smooth
function of the indicated arguments, the consistency condition for equation (%
\ref{4}) imply that 
\begin{equation}
h_{ij}=h_{ji}\,,\;\;\;\;\frac{\partial h_{ij}}{\partial\dot{q}^{k}}=\frac{%
\partial h_{kj}}{\partial\dot{q}^{i}}\,.  \label{6}
\end{equation}
If (\ref{6}) does hold one can solve equation (\ref{4}). To this end, we
remind that the general solution of the equation $\partial f/\partial
q^{i}=g_{i}$, provided the vector $g_{i}$ is a gradient, is 
\begin{equation*}
f(q)=\int_{0}^{1}ds\,\, q^{i}g_{i}(sq)\,+c\,,
\end{equation*}
where $c$ is a constant. Taking the above fact into account, we obtain for $%
L $ (we do not consider global problems which can arise from non-trivial
topology of the configuration space) the following representation: 
\begin{equation}
L=K(t,q,\dot{q})+l_{i}(t,q)\dot{q}^{i}+l_{0}(t,q)\,,  \label{7}
\end{equation}
where 
\begin{equation}
K(t,q,\dot{q})=\overset{1}{\underset{0}{\int}}da\,\dot{q}^{j}\left[\overset{1%
}{\underset{0}{\int}}db\,\dot{q}_{1}^{i}h_{ij}(t,q,b\dot{q}_{1})\right]_{%
\dot{q}_{1}=a\dot{q}}  \label{7a}
\end{equation}
and $l_{0}(t,q)$, $l_{i}(t,q)$ are some functions of the indicated
arguments. To find these functions, we use equation (\ref{5}). Substituting (%
\ref{7}) into (\ref{5}), we get%
\begin{equation}
\frac{\partial K}{\partial q^{i}}-\frac{\partial^{2}K}{\partial\dot{q}%
^{i}\partial t}-\frac{\partial^{2}K}{\partial\dot{q}^{i}\partial q^{j}}\dot{q%
}^{j}+\left(\frac{\partial l_{j}}{\partial q^{i}}-\frac{\partial l_{i}}{%
\partial q^{j}}\right)\dot{q}^{j}-\frac{\partial l_{i}}{\partial t}+\frac{%
\partial l_{0}}{\partial q^{i}}=h_{ij}f^{j}~.  \label{8}
\end{equation}
Differentiating this equation over $\dot{q}^{k},$ we obtain:%
\begin{equation}
\frac{\partial l_{k}}{\partial q^{i}}-\frac{\partial l_{i}}{\partial q^{k}}%
=L_{ik}~,  \label{9q}
\end{equation}
where%
\begin{equation}
L_{ik}=\frac{\partial^{2}K}{\partial\dot{q}^{i}\partial q^{j}}-\frac{%
\partial^{2}K}{\partial\dot{q}^{j}\partial q^{i}}+\frac{\partial h_{ik}}{%
\partial t}+\dot{q}^{j}\frac{\partial h_{ik}}{\partial q^{j}}+\frac{\partial%
}{\partial\dot{q}^{k}}\left(h_{ij}f^{j}\right)\ .  \label{10}
\end{equation}
This equation is a differential equation for $l_{i}$. The consistency
conditions of equation (\ref{9q}) imply that, first of all, the symmetric
part of $L_{ik}$ is zero, which can be written as 
\begin{equation}
\hat{D}h_{ik}+\frac{1}{2}\left(h_{ij}\frac{\partial f^{j}}{\partial\dot{q}%
^{k}}+h_{kj}\frac{\partial f^{j}}{\partial\dot{q}^{i}}\right)=0\,,
\label{11}
\end{equation}
where 
\begin{equation*}
\hat{D}=\frac{\partial}{\partial t}+\dot{q}^{j}\frac{\partial}{\partial q^{j}%
}+f^{j}\frac{\partial}{\partial\dot{q}^{j}}\,.
\end{equation*}
Using (\ref{11}), one can rewrite (\ref{10}) as follows%
\begin{equation}
L_{ik}=\frac{\partial^{2}K}{\partial\dot{q}^{i}\partial q^{k}}-\frac{%
\partial^{2}K}{\partial\dot{q}^{k}\partial q^{i}}+A_{ik}~,\ \ A_{ik}=\frac{1%
}{2}\left(h_{ij}\frac{\partial f^{j}}{\partial\dot{q}^{k}}-h_{kj}\frac{%
\partial f^{j}}{\partial\dot{q}^{i}}\right)  \label{12q}
\end{equation}
Next, $L_{ik}$ does not depend on the velocities, i.e., $\partial
L_{ik}/\partial\dot{q}^{l}=0$, which yields%
\begin{equation}
\frac{\partial h_{kl}}{\partial q^{i}}-\frac{\partial h_{il}}{\partial q^{k}}%
=\frac{\partial}{\partial\dot{q}^{l}}A_{ik}\,.  \label{13}
\end{equation}
And, finally, the Jacobi identity 
\begin{equation}
\frac{\partial L_{ik}}{\partial q^{l}}+\frac{\partial L_{kl}}{\partial q^{i}}%
+\frac{\partial L_{li}}{\partial q^{k}}=0\ \Rightarrow\frac{\partial A_{ik}}{%
\partial q^{l}}+\frac{\partial A_{kl}}{\partial q^{i}}+\frac{\partial A_{li}%
}{\partial q^{k}}=0\,.  \label{14}
\end{equation}
Provided $h_{ij}$ obeys equations (\ref{11}), (\ref{13}) and (\ref{14}), $%
l_{i}$ can be found from equation (\ref{9q}). We remind that the general
solution for $l_{i}$ of equation (\ref{9q}) is given by 
\begin{equation}
l_{i}(t,q)=\overset{1}{\underset{0}{\int}}da\, q^{k}L_{ki}(t,aq)+\frac{%
\partial\varphi\left(t,q\right)}{\partial q^{i}}\,,  \label{15}
\end{equation}
where $\varphi(t,q)$ is an arbitrary function.

Now from equation (\ref{8}) we can find $l_{0}$; to this end let us rewrite
it as follows:%
\begin{equation}
\frac{\partial l_{0}}{\partial q^{i}}=m_{i}~,  \label{16}
\end{equation}
where%
\begin{equation}
m_{i}=h_{ij}f^{j}-\frac{\partial K}{\partial q^{i}}+\frac{\partial^{2}K}{%
\partial t\partial\dot{q}^{i}}+\dot{q}^{j}\frac{\partial^{2}K}{\partial
q^{j}\partial\dot{q}^{i}}-\dot{q}^{j}L_{ij}+\frac{\partial l_{i}}{\partial t}%
~.  \label{17}
\end{equation}
The consistency conditions of (\ref{16}) imply that, first, $m_{i}$ does not
depend on the velocities, i.e., $\partial m_{i}/\partial\dot{q}^{k}=0$. This
condition is provided by equation (\ref{11}). And second, the vector $m_{i}$
must be a gradient:%
\begin{equation}
\frac{\partial m_{i}}{\partial q^{k}}-\frac{\partial m_{k}}{\partial q^{i}}=%
\frac{\partial A_{ik}}{\partial t}+\dot{q}^{j}\frac{\partial A_{ik}}{%
\partial q^{j}}+\frac{\partial}{\partial q^{k}}(h_{ij}f^{j})-\frac{\partial}{%
\partial q^{i}}(h_{kj}f^{j})=0\,.  \label{18}
\end{equation}
Taking into account (\ref{6}), (\ref{11}) and (\ref{13}), one gets from (\ref%
{18}) the following algebraic condition: 
\begin{equation}
h_{ij}B_{k}^{j}-h_{kj}B_{i}^{j}=0\,,  \label{19}
\end{equation}
where 
\begin{equation*}
B_{j}^{i}=\frac{1}{2}\frac{\partial f^{i}}{\partial\dot{q}^{m}}\frac{%
\partial f^{m}}{\partial\dot{q}^{j}}-\hat{D}\frac{\partial f^{i}}{\partial%
\dot{q}^{j}}+2\frac{\partial f^{i}}{\partial q^{j}}\,.
\end{equation*}
If $h_{ij}$ obeys (\ref{19}), then from (\ref{16}) one gets 
\begin{equation}
l_{0}(t,q)=\overset{1}{\underset{0}{\int}}da\, q^{k}m_{k}(t,aq)\,+\frac{%
\partial\varphi\left(t,q\right)}{\partial t}+c(t),  \label{20}
\end{equation}
where $c\left(t\right)$ is an arbitrary function of time.

Thus, we have proved the following statement\textbf{:} \textit{iff for a
given set of second-order ordinary differential equations (\ref{1}) there
exists a non-singular matrix} $h_{ij}\left( t,q,\dot{q}\right) $\textit{that
obeys equations (\ref{6}), (\ref{11}), (\ref{13}), (\ref{14}) and (\ref{19}%
), then this set can be obtained from the variational principle with
Lagrangian (\ref{7}), where the functions} $K\left( t,q,\dot{q}\right) ,\
l_{i}\left( t,q\right) $\textit{\ and} $l_{0}\left( t,q\right) $\textit{\
are defined by (\ref{7a}), (\ref{15}) and (\ref{20}) respectively, and the
functions} $\varphi \left( t,q\right) $\textit{\ and} $c\left( t\right) $%
\textit{\ are arbitrary functions of the indicated arguments.}

The arbitrariness related to the functions $\varphi\left(t,q\right)$ and $%
c\left(t\right)$ enter the Lagrangian (\ref{7}) via the total time
derivative of a function $F,$%
\begin{equation*}
F=\varphi\left(t,q\right)+\int c\left(t\right)dt.
\end{equation*}

Note that an integrating multiplier $h_{ij},$ and as a consequence the
Lagrange function $L$ does exist, but however not for any set of equations (%
\ref{1}). In Section 3 we consider an example of a dynamical system which
does not admit the existence of an integrating multiplier. However, if it
exists, it is not unique \cite{GK}-\cite{Tempesta}, e.g., if the matrix $%
h_{ij}$ is an integrating multiplier for a certain set (\ref{1}), it is easy
to see that the matrix $\acute{h}_{ij}=c$ $h_{ij}$, where $c\neq0$ is a
constant, is an integrating multiplier as well. Therefore, Lagrangian (\ref%
{7}) leading to the set of equations (\ref{1}) is not unique; because for
this set there exist as many inequivalent Lagrangians as integrating
multipliers. Lagrangians corresponding to different integrating multipliers
are known as $s$-equivalent Lagrangians.

In the one dimensional case $\ddot{q}-f(t,q,\dot{q})=0,$ an integrating
multiplier is a non-vanishing function $h(t,q,\dot{q})$ that obeys the
equation%
\begin{equation}
\frac{\partial h}{\partial t}+\dot{q}\frac{\partial h}{\partial q}+\frac{%
\partial}{\partial\dot{q}}\left(fh\right)=0\,.  \label{21}
\end{equation}
This is a first-order partial differential equation which obviously has a
solution for any $f$ and initial condition $h\left(t=0,q,\dot{q}%
\right)=h_{0}\left(q,\dot{q}\right)$. As we can see, an answer to the
question whether there exists a solution of the inverse problem of the
calculus of variations depends on the number of degrees of freedom $n$. For $%
n=1$ the answer is always positive, and there exist as many inequivalent
Lagrangians as functions $h_{0}\left(q,\dot{q}\right)$ of two variables. For 
$n\geq2$ the answer is generally negative.

\subsection{Examples}

In this section we consider the possibility of constructing an action
principle for some examples of dynamical systems. First of all, let us
consider dissipative systems. Suppose we have an ideal system with
Lagrangian 
\begin{equation}
L_{0}=\frac{\dot{q}^{2}}{2}+V(q),\,\,\,\;\;\; q=\{q^{i}\},\;\;\;\; i=1...n.
\label{22}
\end{equation}
Let us consider the case when besides the potential conservative force $%
F^{i}=\frac{\partial V}{\partial q^{i}}$ there exist a friction force 
\begin{equation}
F_{fric}^{i}=\alpha\dot{q}^{i},  \label{23}
\end{equation}
where $\alpha$ is a phenomenological friction coefficient which in general
can depend on time. The equations of motion for such a system have the form%
\begin{equation}
\ddot{q}^{i}=\frac{\partial V}{\partial q^{i}}+\alpha\dot{q}^{i}.  \label{24}
\end{equation}
These equations are non-Lagrangian, but for this set it is possible to find
an integrating multiplier. In the simplest case, when it does not depend on
coordinates and velocities, it has the form%
\begin{equation}
h_{ij}=e^{-2\int\alpha dt}h_{ij}^{0},  \label{25}
\end{equation}
where $h_{ij}^{0}$ is an arbitrary, symmetric, nonsingular, constant matrix
commuting with the matrix $V_{ij}=\partial^{2}V/\partial q^{i}\partial
q^{j}. $ Using the statement of the previous section, we obtain the
following Lagrangian: 
\begin{equation}
L=\frac{1}{2}\dot{q}^{i}h_{ij}\dot{q}^{j}+\overset{1}{\underset{0}{\int}}%
q^{i}h_{ij}\frac{\partial V(s\vec{q})}{\partial q^{j}}ds.  \label{26}
\end{equation}
If one sets $h_{ij}^{0}=\delta_{ij}$, Lagrangian (\ref{26}) can be rewritten
as 
\begin{equation}
L=e^{-2\int\alpha dt}L_{0}.  \label{27}
\end{equation}
Note that once the friction coefficient goes to zero, Lagrangian (\ref{27})
transforms into the initial Lagrangian (\ref{22}).

Let us now consider the case when the potential in the initial Lagrangian is
linear in velocities. For simplicity we consider the two-dimensional case%
\begin{equation}
L_{0}=\frac{1}{2}\left(\dot{x}^{2}+\dot{y}^{2}+\beta(\dot{x}y-\dot{y}%
x)\right),  \label{28}
\end{equation}
Let us consider this system in the presence of the dissipative force (\ref%
{26}). The equations of motion will have the form:%
\begin{equation}
\begin{array}{l}
\ddot{x}=\alpha\dot{x}-\beta\dot{y}\,, \\[3mm] 
\ddot{y}=\beta\dot{x}+\alpha\dot{y}\,.%
\end{array}
\label{29}
\end{equation}
As was shown in \cite{kup}, this system describes a moving charged particle
in a uniform magnetic field with radiation friction. In this case,%
\begin{equation*}
B_{j}^{i}=\left(%
\begin{array}{cc}
\alpha^{2}-\beta^{2} & -2\alpha\beta \\ 
2\alpha\beta & \alpha^{2}-\beta^{2}%
\end{array}%
\right)\,
\end{equation*}
and from equation (\ref{19}) one immediately gets%
\begin{equation}
tr\left(h_{ij}\right)=h_{11}+h_{22}=0\,.  \label{30}
\end{equation}
It is then easy to find that the general solution of the equations (\ref{6}%
), (\ref{11}), (\ref{13}), (\ref{14}) is defined by an arbitrary function $%
\phi\left(\zeta,\eta\right)$ and has the form%
\begin{equation}
h_{ij}=\left(%
\begin{array}{cc}
F+\bar{F} & i\left(F-\bar{F}\right) \\ 
i\left(F-\bar{F}\right) & -\left(F+\bar{F}\right)%
\end{array}%
\right)\,,  \label{31}
\end{equation}
where $F=\phi\left(\dot{\xi}e^{-\gamma t},\dot{\xi}-\alpha\xi\right)e^{-%
\gamma t},\ \xi=x+iy,\ \gamma=\alpha+i\beta$ and the bar denotes a complex
conjugation.

The simplest real solution can be found if we put $\phi=1/\zeta$. We have%
\begin{equation}
h_{ij}=\frac{2}{\dot{x}^{2}+\dot{y}^{2}}\left(%
\begin{array}{cc}
\dot{x} & \dot{y} \\ 
\dot{y} & -\dot{x}%
\end{array}%
\right)~.  \label{32}
\end{equation}
Using the formulas (\ref{7}), (\ref{7a}), (\ref{15}) and (\ref{20}) we find
the following Lagrangian:%
\begin{equation}
L=\frac{1}{2}\dot{x}\ln\left(\dot{x}^{2}+\dot{y}^{2}\right)+\dot{y}%
\arctan\left(\frac{\dot{x}}{\dot{y}}\right)+\alpha x-\beta y~.  \label{33}
\end{equation}
The corresponding Euler-Lagrange equations%
\begin{equation}
\frac{\ddot{x}\dot{x}+\ddot{y}\dot{y}}{\dot{x}^{2}+\dot{y}^{2}}=\alpha,\ \ 
\frac{\ddot{x}\dot{x}-\ddot{y}\dot{y}}{\dot{x}^{2}+\dot{y}^{2}}=\beta
\label{34}
\end{equation}
are equivalent to the initial ones (\ref{29}), with the exception of the
point $\dot{x}=\dot{y}=0$. Thus, we can see that in this case the inverse
problem of the calculus of variations is solvable. Unfortunately, neither
Lagrangian (\ref{33}) nor any other Lagrangian constructed by the matrix (%
\ref{31}) in the limit $\alpha\rightarrow0$ transforms into the initial
Lagrangian (\ref{28}), modulo a total time derivative. This is because,
according to the algebraic condition (\ref{30}), the trace of the Hessian
matrix of any Lagrangian for the set of equations (\ref{29}) must be equal
to zero and this property holds true after the limit $\alpha\rightarrow0$ is
taken. On the other hand, the trace of the Hessian matrix of the Lagrangian $%
L_{0}$ in (\ref{28}) is equal to $2$. This contradiction proves the
statement.

Finally we consider the example of a dynamical system for which an
integrating multiplier and, consequently, the possibility of the Lagrangian
description does not exist. Duglas \cite{Duglas} showed that the set of
second-order equations 
\begin{eqnarray*}
\ddot{x}+\dot{y} & = & 0, \\
\ddot{y}+y & = & 0
\end{eqnarray*}
does not admit an integrating multiplier.Let us prove this. To this end, let
us assume the opposite, namely, let there be a non-degenerate matrix $h_{ij}$
that obeys equations (\ref{6}), (\ref{11}), (\ref{13}), (\ref{14}) and (\ref%
{19}). Then from the algebraic equation (\ref{19}) it follows that $h_{ij}$
must be diagonal ($h_{12}=h_{21}=0$), since in this case 
\begin{equation*}
B_{j}^{i}=\left(%
\begin{array}{cc}
0 & 0 \\ 
0 & 2%
\end{array}%
\right)\,.
\end{equation*}
Then, from condition (\ref{11}) we immediately obtain $h_{11}=0$, and arrive
at a contradiction, $\det h_{ij}=0.$

Thus, we can see that an action functional in second-order formalism cannot
be constructed for some sets of differential equations. Nevertheless, as we
show in the following section, it is always possible to construct an action
principle for the equivalent set of first-order equations.

\section{Action principle in the first-order form}

Let a system with $n$ degrees of freedom be described by a set of $n$
non-Lagrangian second-order differential equations of motion. To construct
an action principle, we replace these equations (which is always possible to
do by introducing $n$ additional variables, e.g. $p_{i}=\dot{q}^{i}$) by an
equivalent set of $2n$ first-order differential equations, solvable with
respect to time derivatives. Suppose such a set has the form 
\begin{eqnarray}
\dot{x}^{\alpha} & = & f^{\alpha}(t,x)\,,\;  \label{1b} \\
x^{\alpha} & = & (q^{i},p_{i})~,\ \ \alpha=1,..,2n\,,  \notag
\end{eqnarray}
where $f^{\alpha}(t,x)$ are some functions of the indicated arguments and by
dots above we denote time derivatives of coordinates. Let us construct an
action principle for this set. If (\ref{1b}) cannot be directly identified
with Euler-Lagrange equations, then one can find an integrating multiplier,
i.e. a nonsingular matrix $\Omega$\footnote{%
We denote by $\Omega$ the integrating multiplier for the first-order
equations.} which reduces the initial set of differential equations (\ref{1b}%
) to a variational derivative:%
\begin{equation}
g_{\alpha}\left[t\right]=\Omega_{\alpha\beta}\left(\dot{x}%
^{\beta}-f^{\beta}(t,x)\right)=\frac{\delta S}{\delta x^{\alpha}}=0\,.
\label{a1}
\end{equation}
Since $g_{\alpha}\left[t\right]$ is a variational derivative it must obey
the Helmholtz criterion \cite{H}%
\begin{equation}
\frac{\delta g_{\alpha}\left[t\right]}{\delta x^{\beta}\left(s\right)}=\frac{%
\delta g_{\beta}\left[s\right]}{\delta x^{\alpha}\left(t\right)}\,.
\label{a2}
\end{equation}
We will use this condition to find an integrating multiplier $\Omega.$ In
the general case one can assume that $\Omega$ depends on time $t$,
coordinates $x^{\alpha}$ and time derivatives up to order $m$ ($m$ is a
natural number), i.e. $g_{\alpha}\left[t\right]=g_{\alpha}\left(t,x,...,x^{%
\left(m\right)}\right)$. Having in mind this form of $g_{\alpha}$ one
rewrites (\ref{a2}) as%
\begin{equation}
\sum_{i=0}^{m}\frac{\partial g_{\alpha}\left[t\right]}{\partial
x^{\beta\left(i\right)}}\delta^{\left(i\right)}\left(t-s\right)=%
\sum_{j=0}^{m}\frac{\partial g_{\beta}\left[s\right]}{\partial
x^{\alpha\left(j\right)}}\delta^{\left(j\right)}\left(s-t\right)\,,
\label{a3}
\end{equation}
since%
\begin{equation*}
\frac{\delta}{\delta x^{\beta}\left(s\right)}x^{\alpha\left(k\right)}\left(t%
\right)=\left(\frac{d}{dt}\right)^{k}\frac{\delta x^{\alpha}\left(t\right)}{%
\delta x^{\beta}\left(s\right)}=\delta_{\beta}^{\alpha}\delta^{\left(k%
\right)}\left(t-s\right)\,,\;\; k=0,1,...\,.
\end{equation*}
Differentiating the identity $f\left(t\right)\delta\left(t-s\right)=f\left(s%
\right)\delta\left(s-t\right)$ over $t$ one finds%
\begin{equation}
f\left(s\right)\delta^{\left(k\right)}\left(s-t\right)=\left(-1\right)^{k}%
\sum_{l=0}^{k}C_{l}^{k}f^{\left(l\right)}\left(t\right)\delta^{\left(k-l%
\right)}\left(t-s\right)\,,\;\; C_{l}^{k}=\frac{k!}{(k-l)!}\,.  \label{a4}
\end{equation}
Using (\ref{a4}) we rewrite (\ref{a2}) as%
\begin{equation}
\sum_{i=0}^{m}\frac{\partial g_{\alpha}\left[t\right]}{\partial
x^{\beta\left(i\right)}}\delta^{\left(i\right)}\left(t-s\right)=%
\sum_{j=0}^{m}\left(-1\right)^{j}\sum_{l=0}^{j}C_{l}^{j}\left[\left(\frac{d}{%
dt}\right)^{l}\frac{\partial g_{\beta}\left[t\right]}{\partial
x^{\alpha\left(j\right)}}\right]\delta^{\left(j-l\right)}\left(t-s\right)\,.
\label{a5}
\end{equation}
Comparing in (\ref{a5}) the coefficient for $\delta^{\left(k\right)}%
\left(t-s\right)$ one gets equations for $g_{\alpha}\left(t,x,...,x^{\left(m%
\right)}\right)$. When $k=0$ we have%
\begin{equation}
\frac{\partial g_{\alpha}}{\partial x^{\beta}}-\sum_{j=0}^{m}\left(-1%
\right)^{j}\left(\frac{d}{dt}\right)^{l}\frac{\partial g_{\beta}}{\partial
x^{\alpha\left(j\right)}}=0\,.  \label{a6}
\end{equation}
Since $g_{\alpha}\left[t\right]$ depends only on derivatives up to order $m$%
, the coefficient of the higher derivative $x^{\alpha\left(2m\right)}$ in
equation (\ref{a6}) must vanish. This coefficient is $\left(-1\right)^{m}%
\partial^{2}g_{\beta}/\partial x^{\alpha\left(m\right)}\partial
x^{\gamma\left(m\right)}$, which means that $g_{\alpha}\left[t\right]$ must
be linear on the derivatives of order $m$, i.e.%
\begin{equation*}
g_{\alpha}\left[t\right]=a_{\alpha\beta}\left(t,x,...,x^{\left(m-1\right)}%
\right)x^{\beta\left(m\right)}+b_{\alpha}\left(t,x,...,x^{\left(m-1\right)}%
\right)\,,
\end{equation*}
where $a_{\alpha\beta}$ and $b_{\alpha}$ are some functions. Since $\Omega$
is a nonsingular matrix, (\ref{a1}) should be a system of first-order
equations, i.e. we have $m=1$ and $\Omega=\Omega\left(t,x\right)$.

Now comparing the coefficient of $\delta^{\left(1\right)}\left(t-s\right)$
in (\ref{a5}), we obtain:%
\begin{equation}
\Omega_{\alpha\beta}=-\Omega_{\beta\alpha}\,.  \label{2b}
\end{equation}
Then from (\ref{a6}) we have%
\begin{equation*}
\partial_{\beta}\left(\Omega_{\alpha\gamma}f^{\gamma}\right)-\partial_{%
\alpha}\left(\Omega_{\beta\gamma}f^{\gamma}\right)+\partial_{t}\Omega_{%
\alpha\beta}+\dot{x}^{\gamma}\left(\partial_{\beta}\Omega_{\alpha\gamma}-%
\partial_{\alpha}\Omega_{\beta\gamma}+\partial_{\gamma}\Omega_{\beta\alpha}%
\right)=0\,.
\end{equation*}
Since $\Omega$ does not depend on $\dot{x}$ one gets the following equations
for $\Omega$: 
\begin{subequations}
\begin{equation}
\partial_{\alpha}\Omega_{\beta\gamma}+\partial_{\beta}\Omega_{\gamma\alpha}+%
\partial_{\gamma}\Omega_{\alpha\beta}=0  \label{3b}
\end{equation}
and 
\end{subequations}
\begin{equation}
\partial_{t}\Omega_{\alpha\beta}+£_{f}\Omega_{\alpha\beta}=0\,,  \label{4b}
\end{equation}
where $£_{f}\Omega_{\alpha\beta}$ is the Lie derivative of $%
\Omega_{\alpha\beta}$ along the vector field $f^{\gamma}$and $%
\partial_{\alpha}=\partial/\,\partial
x^{\alpha},\;\partial_{t}=\partial/\partial t.$ Thus we see that for a set
of first order equations an integrating multiplier is a nonsingular matrix
which depends only on time $t$ and coordinates $x^{\alpha}$, and obeys the
conditions (\ref{2b}), (\ref{3b}) and (\ref{4b}).

Let us analyze equations (\ref{2b})--(\ref{4b}) for the matrix $\Omega
_{\alpha \beta }$ following our work \cite{GK}. It is known that the general
solution $\Omega _{\alpha \beta }$ of equation (\ref{4b}) can be constructed
with the help of a solution of the Cauchy problem for equations (\ref{1b}).
Suppose that such a solution is known:%
\begin{equation}
x^{\alpha }=\varphi ^{\alpha }(t,x_{(0)})\,,\;x_{\left( 0\right) }^{\alpha
}=\varphi ^{\alpha }(0,x_{(0)})  \label{5b}
\end{equation}%
is a solution of equations (\ref{1b}) for any $x_{(0)}=\left( x_{\left(
0\right) }^{\alpha }\right) ,$ and $\chi ^{\alpha }(t,x)$ is the inverse
function with respect to $\varphi ^{\alpha }(t,x_{(0)}),$ i.e.,%
\begin{equation}
x^{\alpha }=\varphi ^{\alpha }(t,x_{(0)})\Longrightarrow x_{\left( 0\right)
}^{\alpha }=\chi ^{\alpha }(t,x)\,,\;x^{\alpha }\equiv \varphi ^{\alpha
}(t,\chi ^{\alpha })\,,\;\,\partial _{\alpha }\chi ^{\gamma }|_{t=0}=\delta
_{\gamma }^{\alpha }\,.  \label{6b}
\end{equation}%
Then 
\begin{equation}
\Omega _{\alpha \beta }(t,x)=\partial _{\alpha }\chi ^{\gamma }\,\Omega
_{\gamma \delta }^{\left( 0\right) }\left( \chi \right) \,\partial _{\beta
}\chi ^{\delta }\,,  \label{7b}
\end{equation}%
where the matrix $\Omega _{\alpha \beta }^{\left( 0\right) }$ is the initial
condition for $\Omega _{\alpha \beta }$, 
\begin{equation*}
\Omega _{\alpha \beta }(t,x)|_{t=0}=\Omega _{\alpha \beta }^{\left( 0\right)
}(x)\,.
\end{equation*}%
One can see that by choosing the matrix $\Omega _{\alpha \beta }^{\left(
0\right) }(x)$ to be nonsingular and subject to the Jacobi identity, we
guarantee the fulfilment of the same conditions for the complete matrix $%
\Omega _{\alpha \beta }(t,x)$, since components of the latter are given by a
change of variables (\ref{7b}).

Thus, we see that for any set of first order equations (\ref{1b}), an
integrating multiplier always exists, i.e. there always exists a Lagrangian $%
L(t,x,\dot{x})$ which has the set of equations%
\begin{equation}
\Omega_{\alpha\beta}\left(\dot{x}^{\beta}-f^{\beta}(t,x)\right)=0  \label{8b}
\end{equation}
as its Euler-Lagrange equations. It is easy to see that this Lagrangian is
linear in the first-order derivative $\dot{x}^{\alpha}$ since equations (\ref%
{8b}) do not contain second-order derivatives, i.e. the corresponding term $%
\partial^{2}L/\partial\dot{x}^{\alpha}\partial\dot{x}^{\beta}$ vanishes. The
general form of this Lagrangian is 
\begin{equation}
L=J_{\alpha}\dot{x}^{\alpha}-H\,,  \label{9b}
\end{equation}
where $J_{\alpha}=J_{\alpha}(t,x)$ and $H=H(t,x)$ are some functions of the
indicated arguments. The Euler--Lagrange equations corresponding to (\ref{9b}%
) are 
\begin{equation}
\frac{\delta S}{\delta x}=\frac{\partial L}{\partial x}-\frac{d}{dt}\frac{%
\partial L}{\partial\dot{x}}=0\Longrightarrow-\partial_{\alpha}H-%
\partial_{t}J_{\alpha}+\left(\partial_{\alpha}J_{\beta}-\partial_{\beta}J_{%
\alpha}\right)\dot{x}^{\beta}=0\,.  \label{10b}
\end{equation}
Comparing equations (\ref{8b}) and (\ref{10b}) one gets\footnote{%
As a more abstract argument in favour of such a form of $\Omega$ we invoke
the Poincare Lemma, according to which any closed form is locally exact, but
here we do not treate the global problems wich can arise from the
non-trivial topology of $x^{\alpha}$-space.}%
\begin{equation}
\Omega_{\alpha\beta}=\partial_{\alpha}J_{\beta}-\partial_{\beta}J_{\alpha}\,,
\label{11b}
\end{equation}
and 
\begin{equation}
\Omega_{\alpha\beta}f^{\beta}-\partial_{t}J_{\alpha}=\partial_{\alpha}H\,.\,
\label{12b}
\end{equation}
The functions $J_{\alpha}$ and $H$ can be found from conditions (\ref{11b})
and (\ref{12b}) if the matrix $\Omega_{\alpha\beta}$ is given. One can see
that consistency conditions for these equations exactly give us equations (%
\ref{2b})-(\ref{4b}) for an integrating multiplier $\Omega_{\alpha\beta}$.
We recall that the general solution $J_{\alpha}(t,x)$ of the equation (\ref%
{11b}), provided that $\Omega_{\alpha\beta}$ is a given antisymmetric matrix
that obeys the Jacobi identity, is given by 
\begin{equation}
J_{\alpha}(t,x)=\int_{0}^{1}x^{\beta}\Omega_{\beta\alpha}(t,sx)\,
sds+\partial_{\alpha}\varphi(t,x)\,,  \label{13b}
\end{equation}
where $\varphi(x)$ is an arbitrary function. Substituting (\ref{7b}) into (%
\ref{13b}), we obtain 
\begin{equation}
J_{\alpha}(t,y)=\int_{0}^{1}y^{\beta}\left.\left[\partial_{\alpha}\chi^{%
\gamma}\,\Omega_{\gamma\delta}^{\left(0\right)}\left(\chi\right)\,\partial_{%
\beta}\chi^{\delta}\right]\right\vert _{x=sy}\,\,
sds+\partial_{\alpha}\varphi(t,y)\,.  \label{14b}
\end{equation}
Equation (\ref{14b}) describes all the ambiguity (arbitrary symplectic
matrix $\Omega_{\gamma\delta}^{\left(0\right)}$ and arbitrary function $%
\varphi(t,x)$) in constructing the term $J_{\alpha}(t,x)$ of the Lagrange
function (\ref{9b}).

To restore the term $H$ in the Lagrange function (\ref{9b}), we need to
solve the equation (\ref{12b}) with respect to $H.$ To this end, we recall
that the general solution of the equation $\partial_{i}f=g_{i}$, provided a
vector $g_{i}$ is a gradient, is given by 
\begin{equation*}
f(x)=\int_{0}^{1}ds\,\, x^{i}g_{i}(sx)\,+c\,,
\end{equation*}
where $c$ is a constant. Taking the above into account, we obtain for $H$
the following representation: 
\begin{equation}
H(t,x)=\int_{0}^{1}ds\, x^{\beta}\left[\Omega_{\beta\alpha}(t,sx)f^{%
\alpha}(t,sx)-\partial_{t}J_{\beta}(t,sx)\right]+c(t)\,,  \label{15b}
\end{equation}
where $c(t)$ is an arbitrary function of time, and $\Omega_{\beta\alpha}$
and $J_{\beta}$ are given by (\ref{7b}) and (\ref{14b}) respectively. All
the arbitrariness in constructing $H$ is thus due to the arbitrary
symplectic matrix $\Omega_{\gamma\delta}^{\left(0\right)}$ and the arbitrary
functions $\varphi(t,x)$ entering the expressions for $\Omega_{\beta\alpha}$
and $J_{\beta}$, and the arbitrary functions $c(t).$

We can see that there exists a family of Lagrangians (\ref{9b}) which lead
to the same equations of motion (\ref{1b}). It is easy to see that actions
with the same $\Omega_{\gamma\delta}^{\left(0\right)}$ but different
functions $\varphi(t,x)$ and $c(t)$ differ by a total time derivative (we
call such a difference trivial). A difference in Lagrange functions related
to different choice of symplectic matrices $\Omega_{\alpha\beta}^{\left(0%
\right)}$ is not trivial. The corresponding Lagrangians are known as $s$%
-equivalent Lagrangians.

As an example, let us consider a theory with equations of motion of the form%
\footnote{%
Here we use matrix notation, $x=\left( x^{\alpha }\right) ,\;A(t)=\left(
A(t)_{\beta }^{\alpha }\right) ,\;j(t)=\left( j(t)^{\alpha }\right)
,\;\alpha ,\beta =1,..,2n$.
\par
{}} 
\begin{equation}
\dot{x}=A(t)x+j(t)\,.  \label{i10}
\end{equation}%
An action principle for such a theory can be constructed (see \cite{GK})
following the above described manner.

The solution of the Cauchy problem for the equations (\ref{i10}) reads 
\begin{equation}
x(t)=\Gamma(t)x_{(0)}+\gamma(t)\,,  \label{i11}
\end{equation}
where the matrix $\Gamma(t)$ is the fundamental solution of (\ref{i10}),
i.e., 
\begin{equation}
\dot{\Gamma}=A\Gamma\,,\;\Gamma(0)=1\,,  \label{i12}
\end{equation}
and $\gamma(t)$ is a partial solution of (\ref{i10}). Then following (\ref%
{7b}), we construct the matrix $\Omega$\footnote{%
For simplicity we choose the matrix $\Omega^{\left(0\right)}$ to be a
constant nonsingular antisymmetric matrix.}, 
\begin{equation}
\Omega=\Lambda^{T}\Omega^{\left(0\right)}\Lambda\,,\;\;\Lambda=\Gamma^{-1}\,.
\label{i13}
\end{equation}
and find the functions $J$ and $H$ according to (\ref{14b}) and (\ref{15b}), 
\begin{equation}
J=\frac{1}{2}x\Omega\,,\; H=\frac{1}{2}xBx-Cx\,,  \label{i14}
\end{equation}
where 
\begin{equation}
B=\frac{1}{2}\left(\Omega A-A^{T}\Omega\right)\,,\; C=\Omega j\,.
\label{i15}
\end{equation}
Thus, the action functional for the general quadratic theory is 
\begin{equation}
S[x]=\frac{1}{2}\int dt\left(x\Omega\dot{x}-xBx-2Cx\right)\,.  \label{i16}
\end{equation}

In conclusion we note that it is always possible to construct a Lagrangian
action for any set of non-Lagrangian equations in an extended configuration
space following a simple idea first proposed by Bateman \cite{Bateman}. Such
a Lagrangian has the form of a sum of initial equations of motion being
multiplied by the corresponding Lagrangian multipliers and new variables.
The Euler-Lagrange equations for such an action contain besides the initial
equations some new equations of motion for the Lagrange multipliers. In such
an approach one has to think how to interpret the new variables already on
the classical level. Additional difficulties (indefinite metric) can appear
in course of the quantization \cite{Dekker}-\cite{Blasone1}.

\begin{acknowledgement}
Gitman is grateful to the Brazilian foundations FAPESP and CNPq for
permanent support; Kupriyanov thanks FAPESP for support.
\end{acknowledgement}

\end{document}